\documentclass[twocolumn]{aastex63}

\def\rmit#1{{\it #1}}              
\def\specchar#1{{\sc #1}}
\def\FeI{\mbox{Fe\,\specchar{i}}}

\def\SiI{\mbox{Si\,\specchar{i}}}

\def\HeI{\mbox{He\,\specchar{i}}}
\def\CaII{\mbox{Ca\,\specchar{ii}}} 
\def\Ca{\mbox{Ca}} 
\def\eg{\rmit{e.g.}}

\newcommand{\GG}[1]{}

\received{XX XX, XX}
\revised{XX XX, XX}
\accepted{XX XX, XX}

\submitjournal{ApJ}

\shorttitle{Chromospheric magnetic field fluctuations in umbrae}
\shortauthors{Felipe et al.}

\graphicspath{{./}{figures/}}

\usepackage{multirow}

\begin{document}


\title{Limitations of the \CaII\ 8542 \AA\ line for the determination of magnetic field oscillations}

\correspondingauthor{Tobias Felipe}
\email{tobias@iac.es}

\author[0000-0003-1732-6632]{Tobias Felipe}
\affiliation{Instituto de Astrof\'{\i}sica de Canarias \\
38205 C/ V\'{\i}a L{\'a}ctea, s/n, La Laguna, Tenerife, Spain}
\affiliation{Departamento de Astrof\'{\i}sica, Universidad de La Laguna \\
38205, La Laguna, Tenerife, Spain} 

\author[0000-0001-9896-4622]{Hector Socas Navarro}
\affiliation{Instituto de Astrof\'{\i}sica de Canarias \\
38205 C/ V\'{\i}a L{\'a}ctea, s/n, La Laguna, Tenerife, Spain}
\affiliation{Departamento de Astrof\'{\i}sica, Universidad de La Laguna \\
38205, La Laguna, Tenerife, Spain} 

\author[0000-0002-9309-3298]{C. R. Sangeetha}
\affiliation{Instituto de Astrof\'{\i}sica de Canarias \\
38205 C/ V\'{\i}a L{\'a}ctea, s/n, La Laguna, Tenerife, Spain}
\affiliation{Departamento de Astrof\'{\i}sica, Universidad de La Laguna \\
38205, La Laguna, Tenerife, Spain}

\author[0000-0003-4446-1696]{Ivan Milic}
\affiliation{Department of Physics, University of Colorado, Boulder, CO 80309, USA}
\affiliation{Laboratory for Atmospheric and Space Physics, University of Colorado, Boulder, CO 80303, USA }
\affiliation{National Solar Observatory, Boulder, CO 80303, USA}

\begin{abstract}
Chromospheric umbral oscillations produce periodic brightenings in the core of some spectral lines, known as umbral flashes. They are also accompanied by fluctuations in velocity, temperature, and, according to several recent works, magnetic field. In this study, we aim to ascertain the accuracy of the magnetic field determined from inversions of the \CaII\ 8542 \AA\ line. We have developed numerical simulations of wave propagation in a sunspot umbra. Synthetic Stokes profiles emerging from the simulated atmosphere were computed and then inverted using the NICOLE code. The atmospheres inferred from the inversions have been compared with the original parameters from the simulations. Our results show that the inferred chromospheric fluctuations in velocity and temperature match the known oscillations from the numerical simulation. In contrast, the vertical magnetic field obtained from the inversions exhibits an oscillatory pattern with a $\sim$300 G peak-to-peak amplitude which is absent in the simulation. We have assessed the error in the inferred parameters by performing numerous inversions with slightly different configurations of the same Stokes profiles. We find that when the atmosphere is approximately at rest, the inversion tends to favor solutions that underestimate the vertical magnetic field strength. On the contrary, during umbral flashes, the values inferred from most of the inversions are concentrated at stronger fields than those from the simulation. Our analysis provides a quantification of the errors associated with the inversions of the \CaII\ 8542 \AA\ line and suggests caution with the interpretation of the inferred magnetic field fluctuations.

\end{abstract}

\keywords{Solar chromosphere --- Sunspots --- Solar atmosphere --- Solar oscillations --- Computational methods --- Observational astronomy}

\section{Introduction} \label{sect:intro}

The study of waves and oscillations in solar active regions has gathered the attention of solar physicists over the last decades. Significant progress has been achieved thanks to the improvement of the observing capabilities and the theoretical modeling \citep[see][for a review]{Khomenko+Collados2015}. Most of the observational works are based on the analysis of velocity and intensity temporal series \citep[\eg,][]{Lites+etal1982, Jess+etal2012a, Tian+etal2014, Cho+etal2015, GilchristMillar+etal2021} since the detection of those fluctuations is straightforward. In contrast, the measurement of magnetic field oscillations poses an observational challenge and their interpretation is associated with intrinsic difficulties.

The first studies addressing magnetic field oscillations focused on the photosphere. Magnetic field oscillations with 5-minute period have been reported in numerous works in different photospheric spectral lines and in different telescopes/instruments \citep[also the references therein]{1998ApJ...497..464L,1998A&A...335L..97R,1999SoPh..187..389B,BellotRubio+etal2000,Khomenko+etal2003}. Observations in \FeI\ lines at 6173.4, 6302, 6843, and 15650 \AA{} are reported in \citet{,1997SoPh..172...69H,1998ApJ...497..464L,1999SoPh..187..389B,BellotRubio+etal2000} respectively. The magnetic oscillations are also observed in the Michelson Doppler Imager Ni I 6768\AA{} line by \citet{BellotRubio+etal2000,1998ApJ...497..464L,1999ApJ...518L.123N,2000SoPh..192..403N,2012SoPh..280..347K}. \citet{2000SoPh..192..403N} has studied and compared magnetic oscillations in three instruments, finding differences in their measurements. Recently, \citep{2021RSPTA.37900175N} have observed magnetic oscillations in data from the Helioseismic Magnetic Imager Fe I 6173~\AA{} line. They interpreted the phase relations between these oscillations and velocity and intensity fluctuations as slow standing or fast standing surface sausage wave modes. In all the above-mentioned studies the amplitudes of oscillations vary from a few Gauss to $\sim100$~G in different magnetic structures on the Sun. Long-period oscillations in the photospheric magnetic field have also been reported in numerous studies \citep[see Table 1 from][for a summary]{2020A&A...635A..64G}.

Despite the large number of works reporting photospheric magnetic field oscillations, their origin is still a matter of debate \citep{Staude2002}. Several authors have suggested that the measured magnetic oscillations are due to opacity effects \citep{BellotRubio+etal2000, Khomenko+etal2003}. In this scenario, the effective formation height of the line changes due to thermodynamics variations associated with the oscillations. If there is a vertical gradient in the magnetic field, the magnetic field probed by the line can fluctuate. \citet{2012SoPh..279..295L} found 12 and 24~hr period oscillations in the magnetic field due to instrumental artifacts. There are also reported observations of cross-talk between the magnetic signal and other atmospheric parameters \citep{1999ApJ...518L.123N} while some state that cross-talk cannot produce the observed periodicity and amplitude in observed oscillations \citep{2003A&A...403..297M}.

Many works have reported the propagation of sunspot photospheric oscillations to upper atmospheric layers \citep[\eg,][]{Lites1984, Centeno+etal2006, Zhao+etal2016}. In their travel, the amplitude of the waves increases due to the drop of the density and they produce dramatic changes in the umbral chromosphere as they develop into shocks. One of the most remarkable manifestations of these shock waves is the generation of umbral flashes \citep{Beckers+Tallant1969, Wittmann1969}. They are periodic, short-lived brightenings of small regions of sunspot umbrae, commonly observed in the core of chromospheric lines. Numerous works have investigated the nature of umbral flashes by analyzing spectropolarimetric observations \citep{SocasNavarro+etal2000a, SocasNavarro+etal2000b, RouppevanderVoort+etal2003, delaCruz-Rodriguez+etal2013, Henriques+etal2017, Anan+etal2019, Bose+etal2019, Henriques+etal2020}. These studies have converged to some common results that confirm the association of umbral flashes with wave propagation, shock formation, and striking temperature enhancements, but also significant discrepancies have arisen.

One aspect of chromospheric oscillations where discrepancies or even contradictory results are more evident is the presence of magnetic field fluctuations. One of the first studies addressing magnetic fluctuations in sunspot chromospheres \citep{delaCruz-Rodriguez+etal2013} found no evidence of magnetic field oscillations from NLTE inversions of \CaII\ 8542 \AA\ umbral observations. However, they did find variations around 200 G in the penumbra associated with running penumbral waves. Later studies based on inversions of the same spectral line have found a reduction of the magnetic field strength during umbral flashes \citep{Henriques+etal2017} or even the opposite result, with umbral flashes exhibiting magnetic field values up to 270 G higher than those inferred for the quiescent chromosphere \citep{Joshi+delaCruzRodriguez2018}. The later work concluded that these magnetic field fluctuations are not consistent with opacity effects since umbral flashes are formed at a higher geometrical height, where the field strength is expected to be lower. \citet{Houston+etal2018} reported fluctuations in the transverse magnetic field up to 200 G. Their results were obtained from inversions of the \HeI\ 10830 \AA, which probes higher chromospheric layers than the \CaII\ 8542 \AA\ employed by the aforementioned works. Recently, \citet{Houston+etal2020} have found support to the interpretation of umbral flashes as magnetohydrodynamic shocks from the examination of magnetic field fluctuations measured in the \CaII\ 8542 \AA\ line.

The discussion from the previous paragraph shows that the \CaII\ 8542 \AA\ line is currently one of the most favored spectral lines for studies of the solar chromosphere, and more specifically for the analysis of chromospheric magnetic field fluctuations. However, it is well-known that this line has limited sensitivity to the magnetic field (its effective Land\'e factor is $\bar{g}=1.10$). In addition, it is optically thick in the solar chromosphere and radiative transfer makes its interpretation complex and non-trivial. The goal of this paper is to evaluate the accuracy of the magnetic field inferred from NLTE inversions of the \CaII\ 8542 \AA\ line alone. With this aim, we have computed synthetic Stokes profiles of the emerging radiation from a numerical simulation of wave propagation in the umbral atmosphere. Then, the atmospheres resulting from the inversions of those profiles have been compared with the original simulated models. This is the third manuscript from a series of studies following this approach. In the previous two papers, we have analyzed the limitations of instruments where a line spectrum is not acquired instantly but scanned through the wavelengths in time \citep{Felipe+etal2018a} and the effect of spectral resolution \citep{Felipe+EstebanPozuelo2019}. The organization of the paper is as follows: Sections \ref{sect:simulations} and \ref{sect:NICOLE} describe the numerical methods, including the development of the simulations and the synthesis and inversions of the line profiles. In Section \ref{sect:evaluation} we compare the results of the inversions with the actually simulated atmospheres. Finally, the conclusions are discussed in Section \ref{sect:conclusions}.

\begin{figure}[ht!]
\plotone{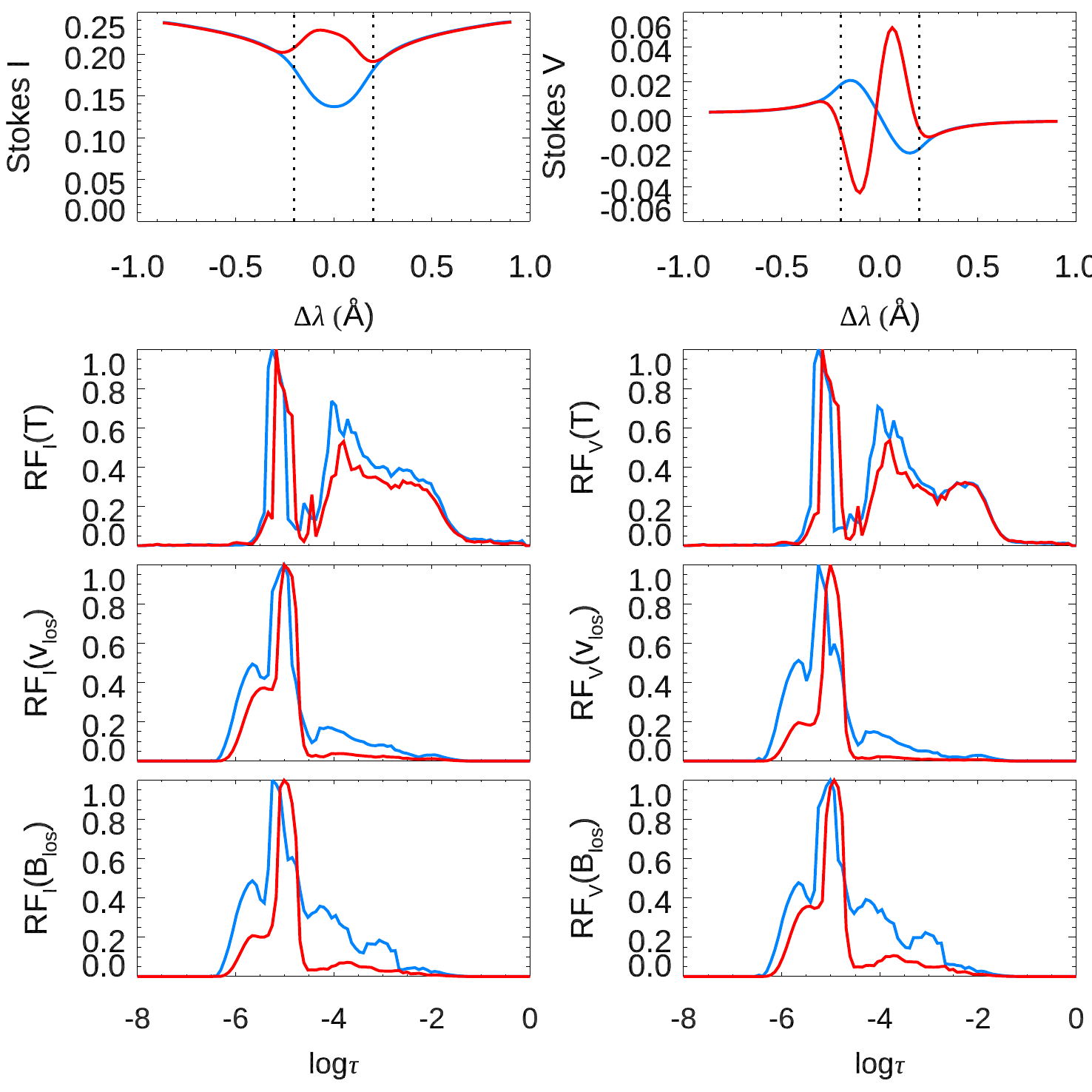}
\caption{Stokes profiles and normalized response functions of the \CaII\ 8542 \AA\ line in the umbral model atmosphere at rest (blue lines) and during an umbral flash (red lines). Top panels show Stokes I (left) and Stokes V (right). Bottom panels illustrate the response functions of the intensity (left column) and Stokes V (right column) to temperature (second row), line-of-sight velocity (third row), and line-of-sight magnetic field (bottom row). Each panel displays the response functions averaged in a spectral range delimited by $\Delta\lambda=\pm0.2$ \AA\ (enclosed by the vertical dotted lines in the top row) as a function of the optical depth.
\label{fig:RFs}}
\end{figure}

\section{Numerical simulations} \label{sect:simulations}

Numerical simulations of wave propagation from the solar interior to the corona in a sunspot umbra were computed with the code MANCHA \citep{Khomenko+Collados2006,Felipe+etal2010a}. This study is based on the analysis of the same simulation developed in \citet{Felipe+etal2021}. The simulations are calculated in a two-dimensional domain but following the 2.5D approximation, which means that the three components of the vectors are maintained. The computational domain spans from $z=-1.14$ Mm to $z=3.50$ Mm, with $z=0$ corresponding to the solar surface (height with optical depth unity at 5000 \AA), and covers 4.8 Mm in the horizontal directions. The vertical and horizontal spatial steps are 10 and 50 km, respectively. 
Waves are excited by a driver located 0.18 Mm below the surface and acting continuously during the whole time of the simulation. This driver was obtained from umbral observations in the \SiI\ 10827 \AA\ line \citep[the observations are described in][]{Felipe+etal2018b}, following \citet{Felipe+etal2011} and \citet{Felipe+Sangeetha2020}. They were acquired with a slit spectrograph and, thus, we only have one spatial dimension available. The driver was imposed along 2.5 Mm of the horizontal dimension in the central part of the computational domain. It reproduces the power spectra measured in the umbral photosphere. The waves generated by the driver are free to propagate in a background umbral model. We employed a model based on that from \citet{Avrett1981}, but the following modifications were included: it was expanded to the solar interior by smoothly merging it with Model S \citep{Christensen-Dalsgaard+etal1996}; the height of the chromospheric temperature increase was shifted to higher layers so the synthesis of the \CaII\ 8542 \AA\ line reproduces observed profiles \citep{Felipe+etal2018a, Felipe+etal2021}; and an isothermal corona was set above the sharp temperature rise from the transition region. The same vertical stratification was imposed in the background atmosphere at all horizontal positions. All locations are permeated by a homogeneous vertical magnetic field with 2000 G field strength.
The output of the simulation has a temporal cadence of 5 s. The synthesis and inversion described in the following sections have been performed on all the time steps during the first 27 min of simulation (321 time steps) at the horizontal positions in a range of 2.3 Mm (46 points). In total, we have analyzed the Stokes profiles produced by 14,766 atmospheric models.

\section{Synthetic observations and analysis} \label{sect:NICOLE}

\subsection{Synthesis of the \CaII\ 8542 \AA\ line} \label{sect:synthesis}

The four Stokes parameters of the \CaII\ 8542 \AA\ line generated by the simulated umbral atmosphere have been synthesized using the NLTE code NICOLE \citep{SocasNavarro+etal2015}. A plane-parallel atmosphere is assumed at each pixel. The \Ca\ atom is modeled including five bound levels and a continuum \citep{delaCruz-Rodriguez+etal2012} and the collisional broadening was estimated following \citet{Anstee+O'Mara1995}. In atmospheres permeated by a magnetic field, like those obtained from the numerical simulations, the polarization induced by Zeeman splitting is computed. The synthesis has been performed assuming that the vertical direction of the simulation coincides with the line-of-sight, that is, that the simulated umbra is at the center of the solar disk. An artificial macroturbulence of 1.8 km s$^{-1}$ was added to broaden the synthetic profiles and obtain \CaII\ 8542 \AA\ intensity profiles with a full width half maximum comparable to that from actual observations. The macroturbulence compensates for the absence of small-scale random motions in the simulation. A similar approach has been followed by several works \citep[\eg,][]{delaCruz-Rodriguez+etal2012, Felipe+etal2018a}. 
The synthesis has been computed at wavelengths in the spectral region within $\pm 880$ m\AA\ from the core of the \CaII\ 8542 \AA\ line. Most of the analyses in this manuscript employ a wavelength sampling of 55 m\AA, which is close to the Nyquist frequency of the CRISP instrument \citep{Scharmer+etal2008} at 8542 \AA. For completeness, we have also evaluated the profiles expected from a slit spectrograph. In Section \ref{sect:DKIST}, we present the results from syntheses with a wavelength step of 18.3 m\AA, which is the spectral sampling at 8542 \AA\ from the Visible Spectro-Polarimeter (ViSP) instrument at the Daniel K. Inouye Solar Telescope \citep[DKIST,][]{Rimmele+etal2020}.
Gaussian noise with an average value of $10^{-3}$ in units of continuum intensity has been added to the spectropolarimetric profiles. This value is comparable to the noise measured in sunspot observations \citep{delaCruz-Rodriguez+etal2013}. Obtaining the signal-to-noise selected for these synthetic observations with the employed temporal cadence (5 s) is beyond the capabilities of state-of-the-art instrumentation. We do not aim to mimic a temporal series acquired from actual observations. Instead, our goal is to analyze Stokes profiles comparable to those obtained from the observations, but using a shorter temporal cadence to better sample the evolution and interpretation of the profiles during umbral flashes.

\begin{figure}[ht!]
\plotone{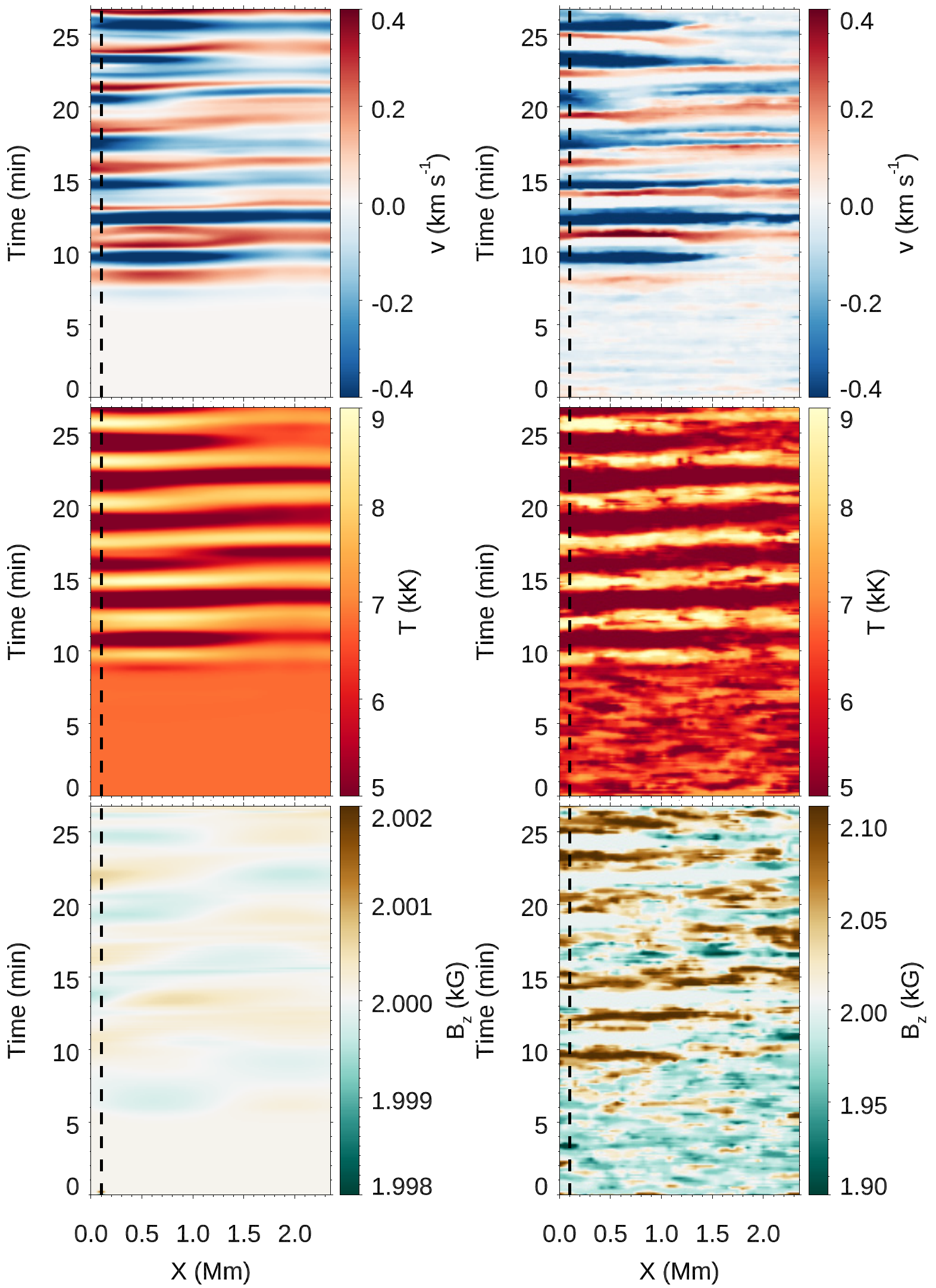}
\caption{Comparison between the chromospheric velocity (top panels), temperature (middle panels), and vertical magnetic field (bottom panels) obtained from the simulation (left panels) and those inferred from the inversion of the synthetic profiles (right panels). The quantities have been averaged in a range of optical depths as indicated in the text. Vertical dashed lines denote the location selected for Figures \ref{fig:comparison} and \ref{fig:comparison_DKIST}.
\label{fig:maps}}
\end{figure}

\subsection{\CaII\ 8542 \AA\ response functions} \label{sect:RFs}

Figure \ref{fig:RFs} illustrates an example of the \CaII\ 8542 \AA\ profiles synthesized from the simulation when the umbral atmosphere is at rest and during an umbral flash. The synthetic profiles exhibit dramatic changes when an umbral flash is taking place, in agreement with observational reports. When the chromosphere is approximately at rest, the core of the line is in absorption and Stokes V shows a regular shape composed of two lobes with opposite sign. During the flash, the core of the line is reversed, showing a slightly blue-shifted peak, whereas Stokes V exhibits a stronger and more complex signal with a flip in the sign of the lobes. This variation in Stokes V is not associated with a change in the magnetic field polarity, but it is produced due to the change of Stokes I from absorption to emission. All these spectral features in the Stokes profiles have been previously found in observational studies \citep[\eg,][]{SocasNavarro+etal2000a, delaCruz-Rodriguez+etal2013}. 
We have computed the response functions of the intensity and Stokes V from the \CaII\ 8542 \AA\ line to temperature, line-of-sight velocity, and line-of-sight magnetic field. The three bottom rows from Figure \ref{fig:RFs} illustrate the response functions from two atmospheric models representing the atmosphere at rest and an umbral flash as a function of $\log\tau$, where $\tau$ is the continuum optical depth at 500 nm. The response functions have been averaged in a spectral region around the core of the line to focus on the wavelengths with stronger contribution from the chromosphere.
The main contribution from all atmospheric parameters is concentrated between $\log\tau\sim-4.7$ and $\log\tau=-5.5$. In all cases, Stokes I and V are sensitive to slightly lower layers (in optical depth scale) during the umbral flash. However, they correspond to higher geometrical heights \citep[see Figure A.1 from][]{Felipe+etal2021}. The comparison of the response functions also shows that the peak with the highest sensitivity to temperature ($\log\tau\sim-5.3$) is slightly higher than the optical depth whose contribution from velocity and magnetic field is maximum ($\log\tau\sim-5.0$).

\subsection{Inversion of the \CaII\ 8542 \AA\ line} \label{sect:inversion}

NICOLE was used in the inversion mode to infer the atmospheric stratification based on the evaluation of the synthetic Stokes profiles generated following Section \ref{sect:synthesis}. The inversion performs successive modifications to an initial guess atmosphere in an iterative process until a good match between its Stokes profiles and those introduced as input is found, as given by the minimization of a $\chi^2$ merit function. During this process, the atmosphere is modified at the location of some selected optical depths known as ``nodes'', and the atmosphere between nodes is obtained from an interpolation. 
For each realization, the inversion is repeated a user-defined number of times with randomized initializations, and the solution with the lowest $\chi^2$ is chosen. The inversions can be speeded up by selecting an acceptable $\chi^2$. In this case, once a sufficiently low $\chi^2$ is achieved, that solution is chosen and the code can proceed to the following case, reducing the total number of computed initializations.   

NICOLE also allows to carry out the inversions by employing several cycles. In each cycle, the solution from the previous cycle is used as the initial guess atmosphere. This way, one can define an inversion strategy where the number of nodes (that is, the number of free parameters) is progressively increased in subsequent cycles. For the results presented in Section \ref{sect:comparison} we chose to invert the profiles using one single cycle with six nodes in temperature, one node in velocity, and three nodes in the vertical magnetic field, and each inversion was repeated up to 25 times until a good solution was found. Only the atmospheres whose Stokes profiles exhibit an excellent agreement with the input profiles are kept. Those cases that did not converge to a low enough $\chi^2$ after the initially planned 25 inversions were repeated setting a higher number of inversion attempts. In Section \ref{sect:spread} we explore the solutions obtained from inversions using different configurations in the distribution of nodes.

\begin{figure*}[ht!]
\plotone{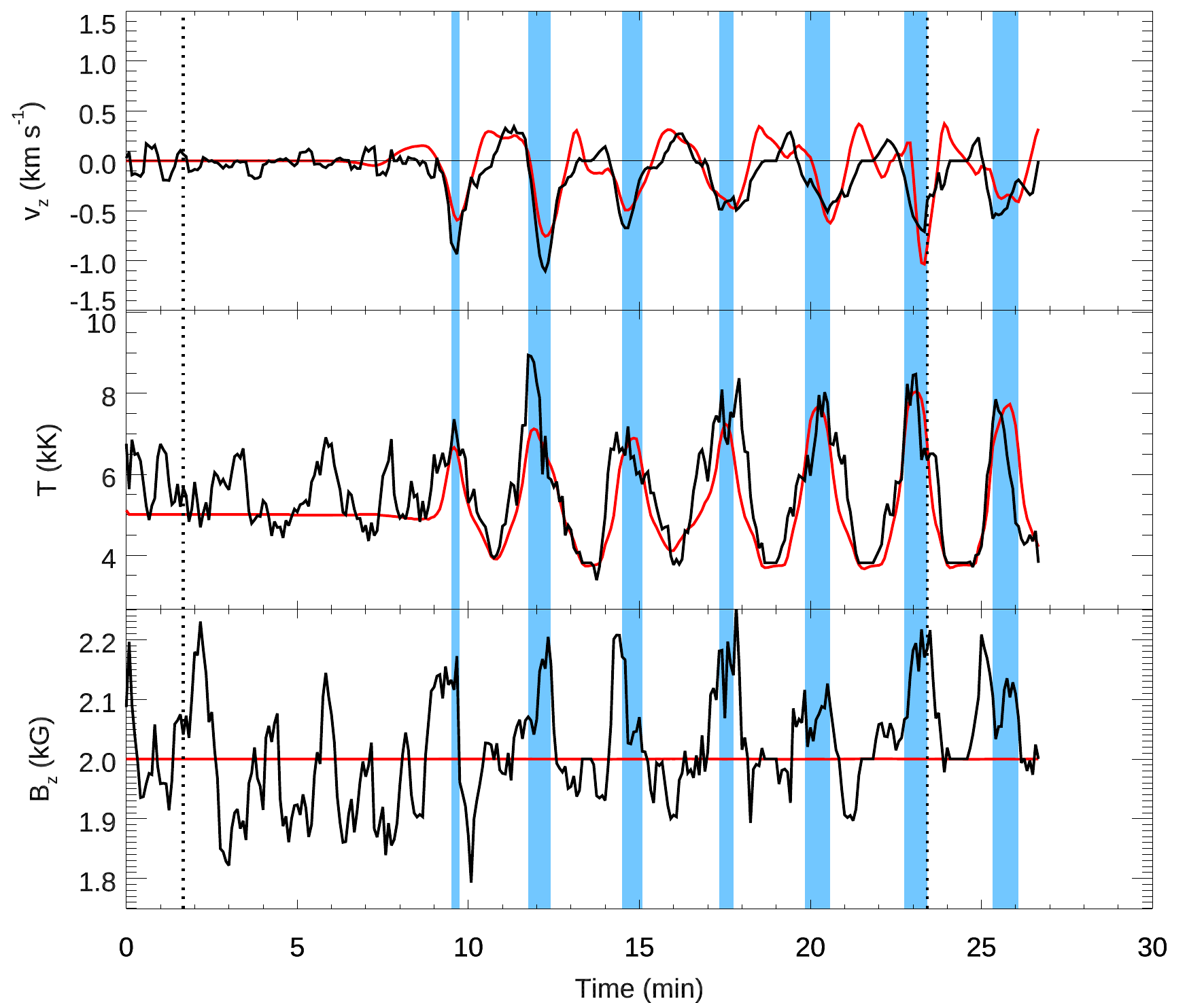}
\caption{Temporal evolution of the chromospheric velocity (top panel), temperature (middle panel), and vertical magnetic field (bottom panel) at a randomly chosen location. In all panels the black line corresponds to the results obtained from the inversions and the red line indicates the actual values from the simulations. In both cases, the quantities have been averaged in a range of optical depths as indicated in the text. Blue-shaded areas denote the times when the core of the \CaII\ 8542 \AA\ line is in emission. Vertical dotted lines indicate the time steps illustrated in Figures \ref{fig:spread_rest}, \ref{fig:spread_UF}, \ref{fig:spread_rest_DKIST}, and \ref{fig:spread_UF_DKIST}.
\label{fig:comparison}}
\end{figure*}

\begin{figure}[ht!]
\epsscale{1.00}
\plotone{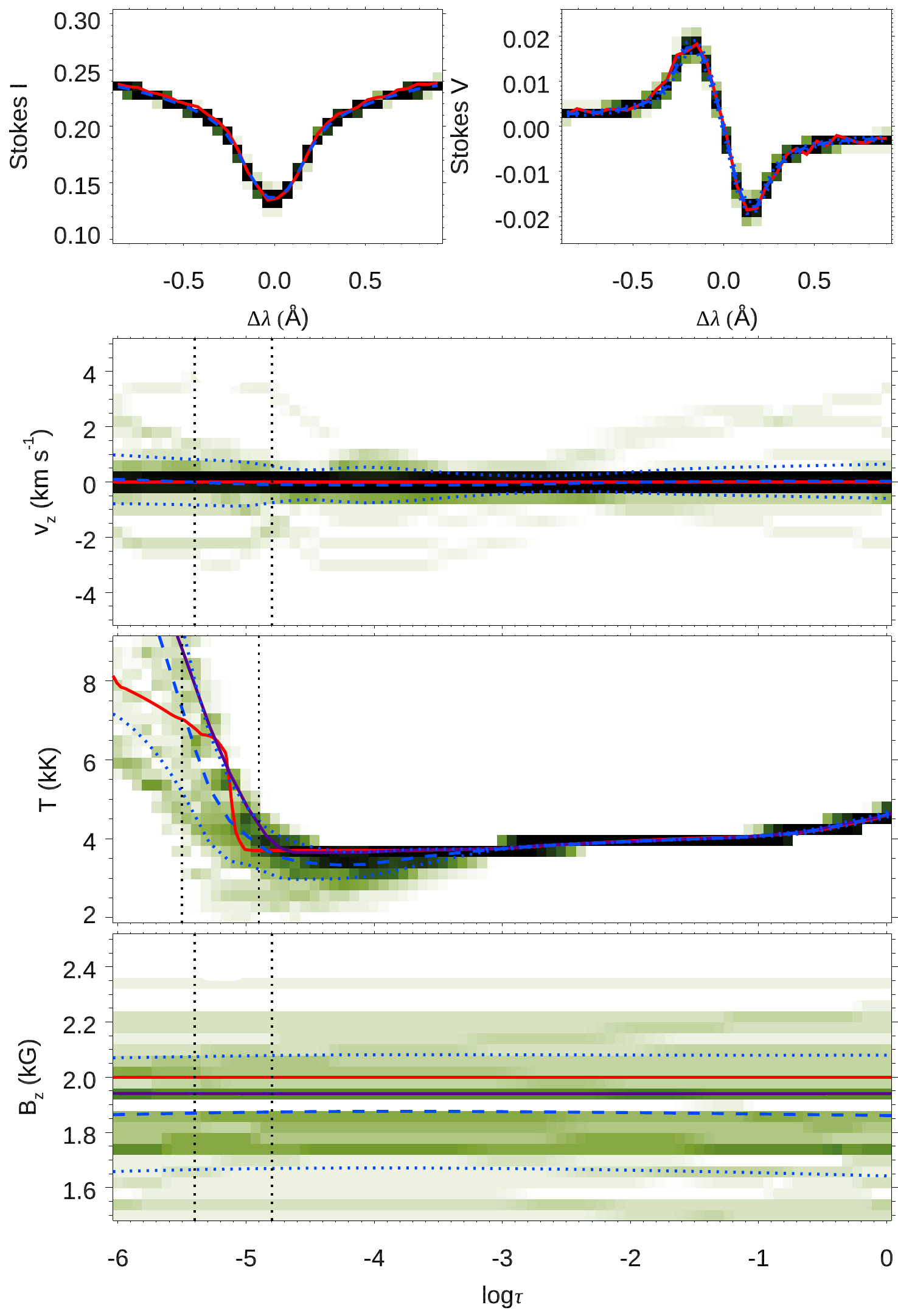}
\caption{Spread of the solutions from a set of inversions of the Stokes profiles from the atmosphere at rest. Top panels illustrate Stokes I (left panel) and Stokes V (right panels). The three lower panels correspond (from top to bottom) to the vertical stratification of the velocity, temperature, and magnetic field as a function of the optical depth. The green density scale indicates the number of inversions for which the code returned a given value, from lower (light green) to higher (dark green) occurrence. Black color indicates an incidence higher than 60\%. Red lines indicate the atmospheric stratification (lower panels) and Stokes profiles (top panels) obtained directly from the numerical simulation and its synthesis, respectively. Blue dashed lines correspond to the average of the inversions, with the error (as given by the standard deviation) showed by the blue dotted lines. In the bottom panels, the violet lines are the atmospheres returned by the best inversion (that with lowest $\chi^2$) and the vertical dotted lines enclose the range of optical depths selected for the average plotted in Fig. \ref{fig:comparison}.
\label{fig:spread_rest}}
\end{figure}

\begin{figure}[ht!]
\epsscale{1.00}
\plotone{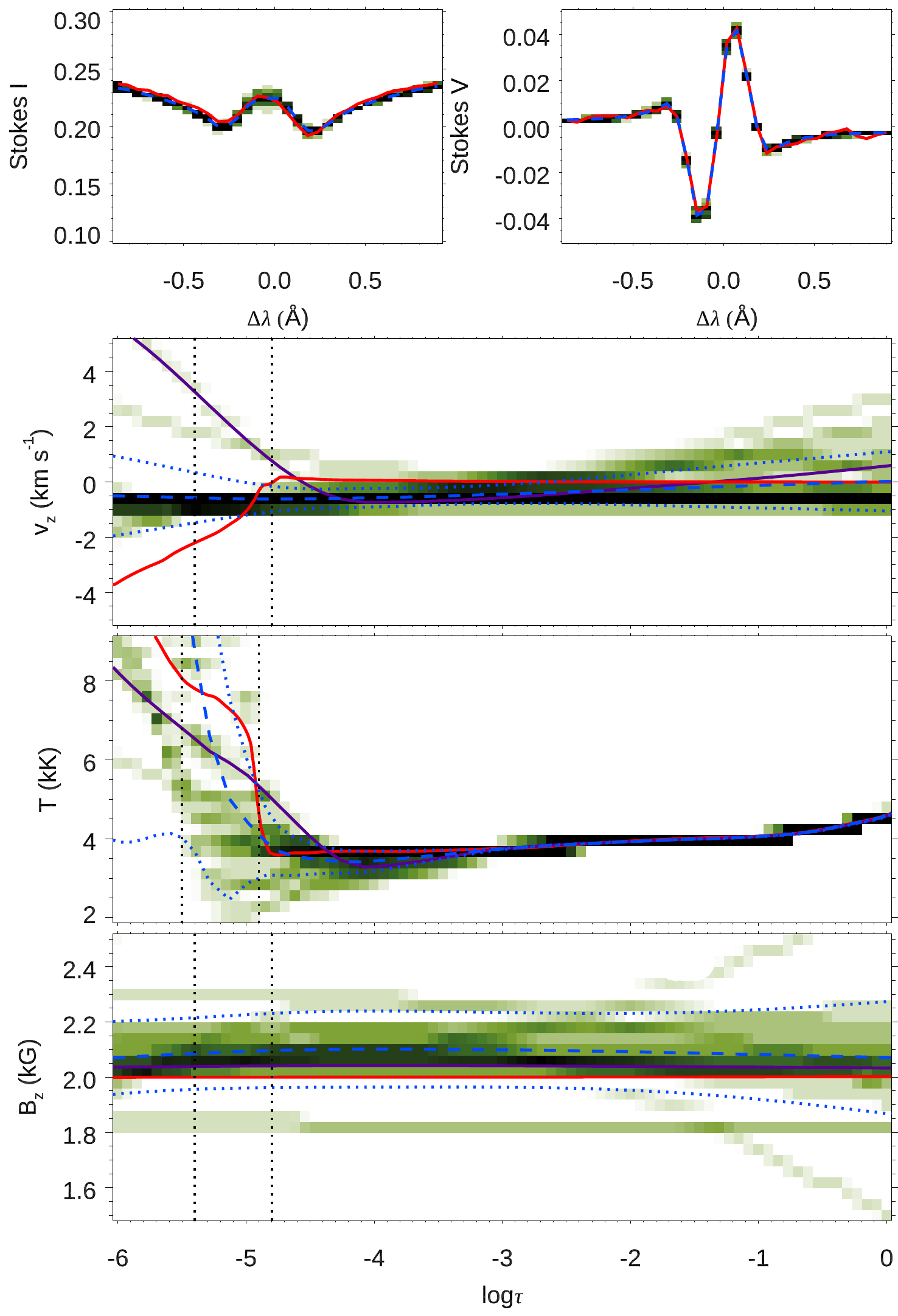}
\caption{Same as Fig. \ref{fig:spread_rest} but for the Stokes profiles of an umbral flash.
\label{fig:spread_UF}}
\end{figure}

\section{Evaluation of the inversions} \label{sect:evaluation}

\subsection{Comparison between simulations and inferred atmospheres} \label{sect:comparison}

Figure \ref{fig:maps} shows the chromospheric temperature, vertical velocity, and vertical magnetic field as a function of horizontal position and time. The actual values (directly obtained from the output of the simulation) are compared with the inferences from the inversion of the synthetic Stokes profiles. In both cases, the quantities have been averaged in the range of atmospheric heights where the response of the line is maximum. Specifically, the velocity and magnetic field have been averaged in $\log\tau=[-4.8,5.4]$, and the temperature in $\log\tau=[-4.9,5.5]$ (see Figure \ref{fig:RFs}). The inverted maps have been smoothed with a boxcar average of width 3$\times$3 (three points in the spatial and temporal dimensions). 
In the numerical simulation, fluctuations are negligible during approximately the first 8 min. This is the time required by the waves to propagate from the subphotospheric layer where they are driven to the atmospheric heights where the line is sensitive. Although during all this time and at all horizontal locations the chromosphere is basically unaltered, the inverted atmospheres exhibit variations in the inferred quantities. Photospheric temperature fluctuations modify the Stokes profiles (Figure \ref{fig:RFs}). The temperature inferred from the inversions is affected by cross talk that causes chromospheric layers to change in order to produce satisfactory fits of atmospheres where only photospheric fluctuations are present. This cross talk between photospheric and chromospheric temperature, possibly due to limitations in the atmosphere representation via nodes, produces the oscillatory pattern found in the chromospheric temperature during the first eight minutes. After those first eight minutes, the velocity and temperature exhibit the well-known three-minute periodicity of the umbral chromospheric oscillations. This periodicity is reasonably captured by the inversion of the Stokes profiles. The inferred chromospheric oscillations show the right phase and amplitude, although discrepancies between the simulated and inverted atmospheres are evident. On the contrary, the inferred magnetic field greatly departs from the actual values. The simulation exhibits magnetic field fluctuations with very low amplitude (around 1 G), whereas the inverted magnetic field oscillates with several hundred Gauss amplitude (note the difference in the color scale displayed in the two bottom panels from Figure \ref{fig:maps}). 


\begin{figure*}[ht!]
\plotone{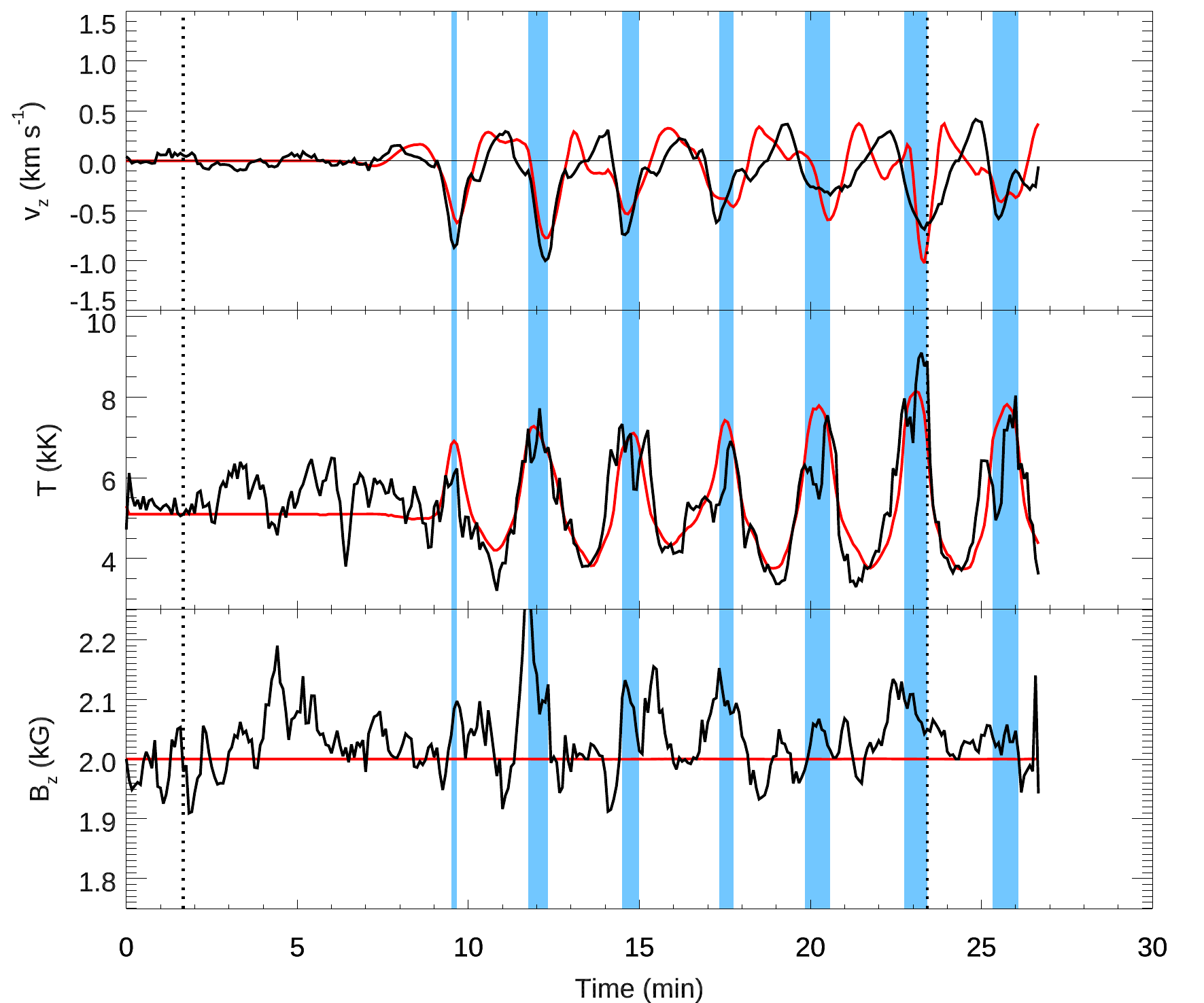}
\caption{Same as Figure \ref{fig:comparison} but obtained from the inversion of Stokes profiles with the finer spectral resolution of a slit spectropolarimeter.
\label{fig:comparison_DKIST}}
\end{figure*}

Figure \ref{fig:comparison} illustrates the comparison between the evolution of the same three quantities in the simulation and the inversions at a randomly chosen location (indicated by vertical dashed lines in Figure \ref{fig:maps}). As previously discussed, the inversions provide a good assessment of the chromospheric thermodynamics, but some differences with the simulations are found. Notably, the maximum of the inferred downward (positive) velocity is systematically lagging the actual velocity. This is in contrast with the upflows (negative velocity), whose timing and amplitude are generally well captured by the inversion. In addition, the inverted temperature exhibits oscillations with peak-to-peak amplitudes of roughly 1000 K during the first 8 min of the simulation, when those fluctuations are absent in the simulation. The bottom panel from Figure \ref{fig:comparison} clearly shows the spurious magnetic field oscillations obtained from the inversion of the \CaII\ 8542 \AA\ line. These oscillations are in phase with those measured in temperature, showing the maximum field strength during umbral flashes (blue-shaded areas). At those times, the inferred magnetic field strength can be even more than 200 G higher than the approximately constant 2000 G vertical magnetic field from the simulation. In contrast, when the synthetic profiles do not exhibit indications of umbral flashes (during the initial 8 min and when the temperature fluctuations are minimum) the inversions show a preference for underestimating the magnetic field strength (see also the prevalence of bluish colors during the first minutes of the simulation in the bottom right panel from Figure \ref{fig:maps}).

\subsection{Spread of the inversion solutions} \label{sect:spread}

We aim to quantify the degeneracy of the inversion problem. We have selected two individual cases, one of them with the umbra at rest and the other representative of the atmosphere during an umbral flash. For each case, we have performed a large number of inversions and we have examined the spread of the solutions. The procedure is as follows. For each set of Stokes profiles, a total of 102 individual inversions have been carried out. These inversions differ in some of the choices that the user generally needs to select when performing NLTE inversions with NICOLE. We have employed six different initial guess atmospheres (by making minor tweaks to the atmosphere employed for the inversions described Section \ref{sect:inversion}) and all of them have been inverted with 17 different node distributions. In all cases, we selected just one cycle. Each of these inversions is repeated 25 times and the solution with the lowest $\chi^2$ is selected. All in all, we have performed 2550 inversions of the same Stokes profiles, and the 102 solutions which best match the input profiles (one of them for each combination of initial guess atmosphere and node distribution) are retained. From these 102 solutions, we only employ for the following analyses those whose $\chi^2$ is below a selected threshold (around 60 inversions for each of the two cases). 

Figure \ref{fig:spread_rest} illustrates the results for the umbral atmosphere approximately at rest, corresponding to a time step from the initial eight minutes of simulation. The red lines show the actual values obtained from the output of the simulation (the stratification of vertical velocity, temperature, and magnetic field shown in the three bottom panels) and from their synthesis (Stokes I and V in the top row). The input profiles/atmosphere are compared with the solutions from the set of inversions. They are illustrated in several ways. The green scale indicates the density of inversion solutions with a certain value. Blue lines represent the average solution from the inversions (dashed lines) and their error as given by the standard deviation (dotted lines). Violet lines in the three bottom panels show the stratification of the solution with the lowest $\chi^2$.  

\begin{figure}[ht!]
\epsscale{1.00}
\plotone{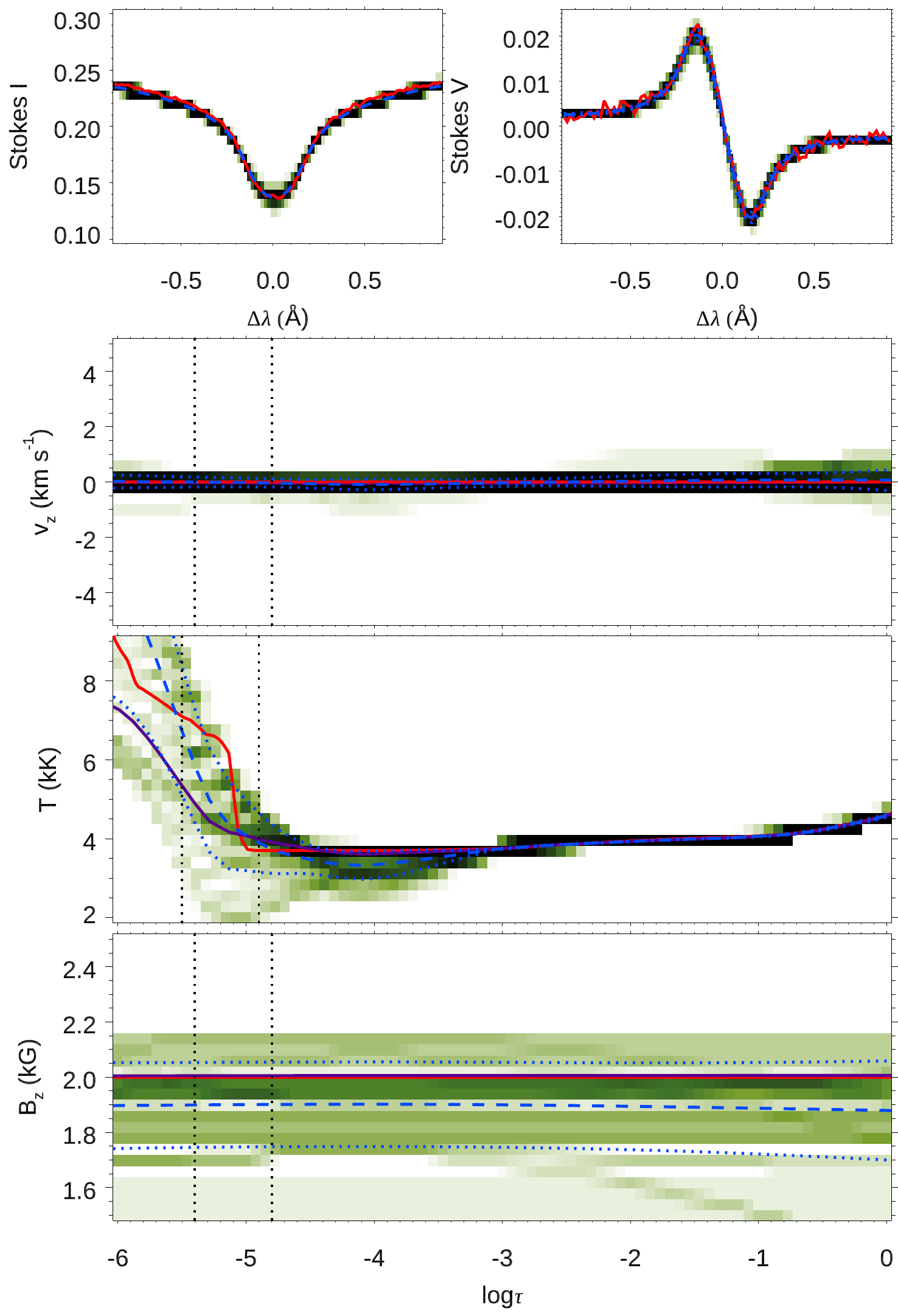}
\caption{Same as Fig. \ref{fig:spread_rest} but with the spectral resolution from a slit spectropolarimeter.
\label{fig:spread_rest_DKIST}}
\end{figure}

\begin{figure}[ht!]
\epsscale{1.00}
\plotone{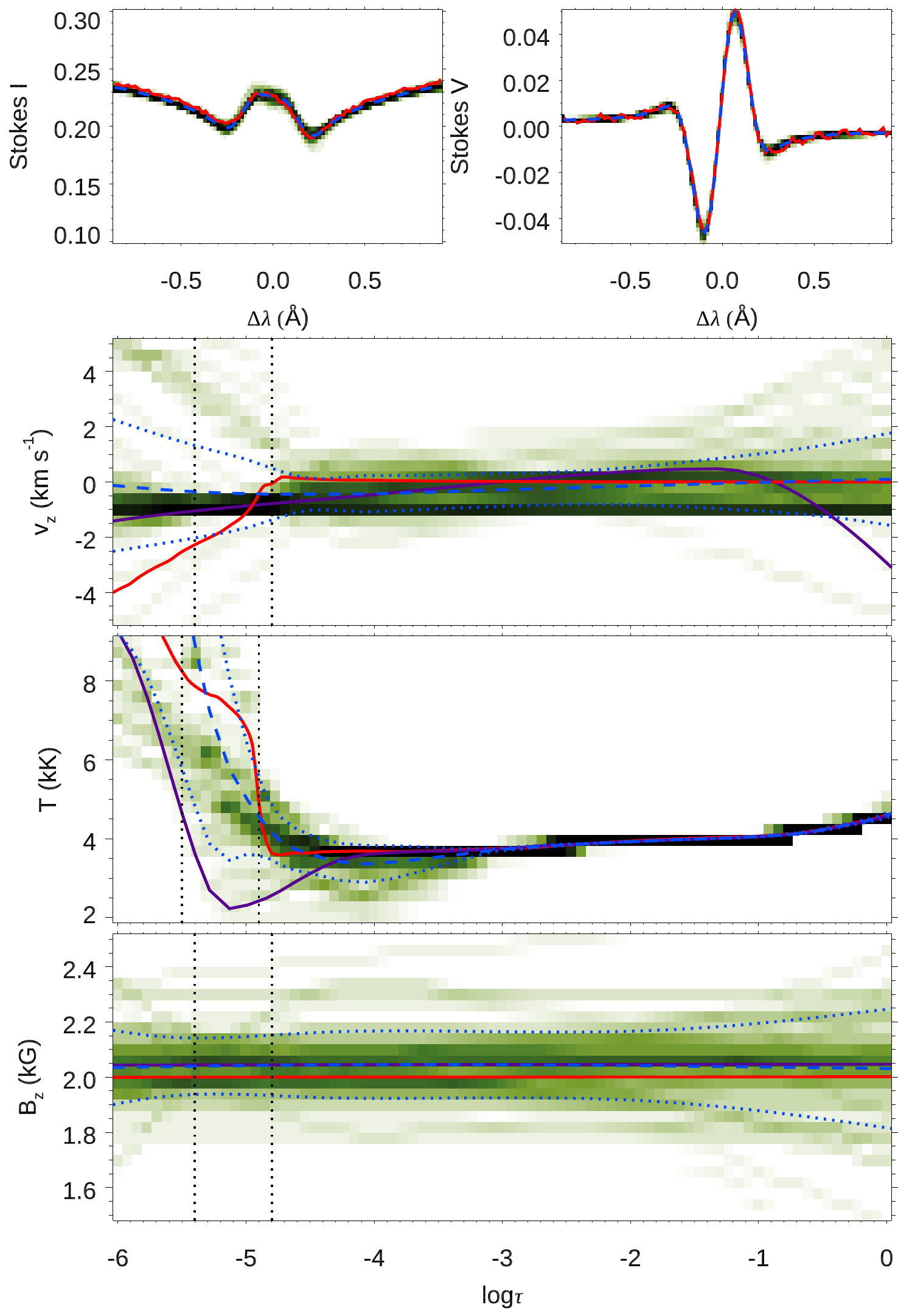}
\caption{Same as Fig. \ref{fig:spread_UF} but with the spectral resolution from a slit spectropolarimeter.
\label{fig:spread_UF_DKIST}}
\end{figure}

The comparison between the input Stokes profiles and the fit obtained from the inversions clearly shows that our analysis is restricted to inversions with exceptional results in terms of quality of the fit. The average inverted Stokes profiles are almost indistinguishable from the input profiles and their dispersion is remarkably low (the standard deviation of Stokes I is not plotted since it is negligible). The velocity of the inferred atmospheres exhibits a very good agreement with the expected value. All the inversions find a velocity close to zero, with an error around 0.5 km s$^{-1}$ at the chromospheric height where the response of the \CaII\ 8542 \AA\ line is maximum. The inferred photospheric temperature also matches the values of the actual atmosphere, but the spread of the solutions increases with height. At the chromospheric heights where the line is sensitive to temperature, the inversions recover a wide range of possible solutions. The temperature rise in the average inferred chromosphere is steeper than in the simulated atmosphere. Inversions cannot probe a discrete height, but they are sensitive to changes taking place at length scales sufficiently long to modify the Stokes profiles. In the case illustrated in Figure \ref{fig:spread_rest}, the average of the inverted chromospheric temperature in the range of optical depths sampled by the spectral line does not depart much from the actual average temperature in the same optical depths. However, there are significant uncertainties in the chromospheric temperature inferred for the atmosphere at rest, as shown in Figure \ref{fig:comparison}. 
The vertical magnetic field obtained from the set of independent inversions of the Stokes profiles from an umbral atmosphere at rest (bottom panel from Figure \ref{fig:spread_rest}) shows discrepancies with the atmosphere generating those profiles. In this atmospheric model, a vertical magnetic field with 2000 G strength is found at all atmospheric layers. In contrast, the magnetic field solutions from the inversions are broadly dispersed. We find that a perfect fit of the Stokes profiles (as those illustrated in top panels from Figure \ref{fig:spread_rest}) can be produced by atmospheric models with strong differences in their chromospheric vertical magnetic field, which ranges from 1470 G to 2346 G. Most of the inversions (73\%) underestimate the actual magnetic field strength, with a higher density of solutions between 1750 and 1950 G. 

Figure \ref{fig:spread_UF} shows the analysis of multiple inversions of the Stokes profiles during an umbral flash. All the selected inversions perfectly capture the line core emission and the reversal in Stokes V (compared to the atmosphere at rest, see Figure \ref{fig:spread_rest}). The chromospheric velocity retrieved by the majority of the solutions is between -0.5 and -1 km s$^{-1}$, which is a good assessment of the average simulated velocity in the range of optical depth sampled by the line. None of the inversions reproduce the steep change in the velocity from $\log\tau=-4.7$ to higher layers, possibly because the line profile is sensitive to an average over the line formation region and not the specific details of how it varies within this region. Complex atmospheres like that could be generated by the inversion code by selecting a higher number of nodes. In the set of configuration files employed for this analysis, there is a prevalence of inversions with a low number of velocity nodes (between one and three). We are only interested in evaluating the inversions at the heights where the line has a strong sensitivity to the atmospheric parameters, and our results show that the chromospheric velocity at $\log\tau\sim-5$ is reasonably probed. Interestingly, some of the inversions with three nodes produce solutions with a strong downflow (positive) velocity. In fact, the best inversion (violet line) exhibits this downflow. \citet{Henriques+etal2017} showed how the inversion of profiles with a blue-shifted feature (like the emission core of the \CaII\ 8542 \AA\ line during the umbral flash) can be fitted by downflowing atmospheres due to radiative transfer artifacts.
During the umbral flash, the chromospheric temperature rise is shifted to a lower height (around $\log\tau\sim-4.8$ in the time step illustrated in Figure \ref{fig:spread_UF}), and the temperature at the formation height of the line is higher. Similarly to the case of the atmosphere at rest, the average of the inversions in the range of optical depths probed by the \CaII\ 8542 \AA\ is consistent with the original temperature. The vertical magnetic field inferred during the umbral flash also exhibits discrepancies with the 2000 G field from the simulation, but some differences with the case of the atmosphere at rest are evident. The inversions of the Stokes profiles of the umbral flash generally overestimate the actual field strength. Also, the dispersion is significantly lower, with the vast majority of the solutions concentrated in the field strength range between 2050 and 2200 G. The higher density of solutions (and the average chromospheric inverted magnetic field) is just below 2100 G.

\subsection{Evaluation of inversions of profiles with high spectral resolution} \label{sect:DKIST}

In Sections \ref{sect:comparison} and \ref{sect:spread} we have evaluated the inversions of Stokes profiles with a 55 m\AA\ sampling, which gives approximately the maximum spectral resolution reachable with an imaging spectropolarimeter like CRISP. Here, we explore the results when a higher spectral resolution is available, as that obtained from slit spectrographs. For this analysis, we have chosen a wavelength step of 18.3 m\AA, which is the expected spectral sampling at 8542 \AA\ from the ViSP instrument at DKIST telescope.

Figure \ref{fig:comparison_DKIST} shows the temporal evolution at the chromosphere from the same location illustrated in Figure \ref{fig:comparison}, but in this case the comparison is done with the results obtained from profiles with higher spectral resolution. Velocity and temperature inferences also exhibit a good characterization of the chromospheric umbral oscillations. The increase in the spectral resolution does not provide a significant improvement in the results. On the contrary, the vertical magnetic field exhibits a better agreement with the values from the simulations. The underestimation of the field strength when the atmosphere is approximately at rest is less common and the lower inferred field strengths are closer to the actual 2000 G value. The inferred field strength peaks are also closer to the real value, except for the strong peak at around 12 min. All in all, the inversion of profiles with this spectral sampling results in the measurement of a spurious oscillation with amplitude $\sim$150 G, whereas in the case of a 55 m\AA\ wavelength step the amplitude of the oscillations is $\sim$300 G. These results are in agreement with \citet{Felipe+EstebanPozuelo2019}, which shows that the errors in the estimation of the magnetic field are reduced with the improvement of the spectral resolutions. In contrast, a finer spectral resolution does not suppose a significant advantage for the evaluation of thermodynamics.
In Figure \ref{fig:spread_rest_DKIST} we examine the solutions from the inversion of Stokes profiles at rest with an 18.3 m\AA\ resolution. Similar to the lower spectral resolution case (Figure \ref{fig:spread_rest}), the estimated velocity and temperature are in good agreement with the original simulated atmospheres. In the case of the velocity, the finer resolution leads to solutions with an approximately zero velocity (in perfect agreement with the simulation) for all the inversions and, thus, a lower standard deviation. The solutions for the vertical magnetic field are also widespread, with most of the inversions providing an underestimated field strength. However, the results show a better agreement with the simulations than the lower spectra resolution analysis since the magnetic field strength in the majority of the inversions is around 1950 G. Remarkably, the best inversion (violet lines) provides a perfect estimation of the vertical magnetic field strength.

When an umbral flash is taking place (Figure \ref{fig:spread_UF_DKIST}), the contribution of the refined spectral resolution to the inference of the magnetic field is also very significant. In this example, the chromospheric vertical magnetic field from most of the inversions is between 1950 and 2150 G, enclosing the original 2000 G field strength from the simulation. The average inferred field strength is 2050 G. The velocity and temperature show similar results to the low spectral resolution analysis (Figure \ref{fig:spread_UF}), although in this case several solutions with a steep velocity increase at the chromosphere (both with downflows and upflows) are found. Surprisingly, the inverted atmosphere producing the best agreement with the input Stokes profiles (violet line) shows the larger departure from the actual chromospheric temperature.

\section{Discussion and conclusions} \label{sect:conclusions}
We have employed inversions to analyze the synthetic \CaII\ 8542 \AA\ Stokes profiles emerging from a known atmosphere, whose temporal evolution was numerically computed by imposing a subphotospheric driver to a background umbral model. We aimed to compare the inferences from the inversions with the original atmosphere to ascertain the sensitivity of NLTE inversions to the fluctuations of the umbral chromosphere produced by wave passages. The simulations reproduce the main properties of the wave phenomena reported from umbral observations. This includes realistic amplitudes (several hundred meters per second in the photosphere, several kilometers per second in the chromosphere), development of shocks, and the change in the dominant period from five minutes in the photosphere to three minutes in the chromosphere \citep[\eg,][]{Kneer+etal1981, Lites1986, Centeno+etal2006, Reznikova+etal2012, Tian+etal2014, KrishnaPrasad+etal2017}. The synthetic \CaII\ 8542 \AA\ Stokes profiles generated by the simulated atmospheres also capture the features found in solar observations, remarkably the appearance of umbral flashes.

The comparison of the inversions and the original atmospheres at the chromospheric heights where the response of the \CaII\ 8542 \AA\ line is maximum reveals a well-founded agreement for the temperature and line-of-sight velocity oscillations and a strong disagreement for the vertical magnetic field. In these simulations, the subphotospheric driver mainly generates a fast magnetoacoustic wave in a high-$\beta$ plasma. Taking into account the vertical magnetic field of the atmosphere and the incidence of this upward propagating wave, at the layer $\beta\sim 1$ it is mostly converted into a slow magnetoacoustic wave \citep{Cally2005}. The behavior of these waves in a low-$\beta$ plasma (like the umbral chromosphere) is similar to a field-aligned acoustic wave. Thus, they barely produce changes in the magnetic field. This is found in the simulations, which show an almost constant chromospheric magnetic field, with the amplitude of the field fluctuations below 1 G. On the contrary, the inversions infer strong magnetic field strength oscillations, whose peak-to-peak amplitude is around 300 G.

We have explored the origin of the magnetic field discrepancy. Several works have suggested that magnetic field fluctuations (measured by photospheric or chromospheric lines) can be produced by opacity effects, due to the variation of the effective formation height of the line in an atmosphere with a vertical gradient in the magnetic field strength due to changes in the thermodynamic parameters during the oscillation \citep[\eg,][]{BellotRubio+etal2000, Khomenko+etal2003, Felipe+etal2014b}. The atmospheric layers where the \CaII\ 8542 \AA\ line is sensitive have been reported to change during flares \citep{Kuridze+etal2018} and umbral flashes \citep{Joshi+delaCruzRodriguez2018}. In these simulations, we have discarded this effect since we chose to permeate the umbral atmosphere with a constant vertical field.  

Inversions commonly do not have a unique solution. Instead, many different atmospheres can leave an indistinguishable imprint in the Stokes profiles, and inversion codes are blind to discriminate among them, particularly when the observations are limited to a single spectral line. To evaluate this degeneracy, we have performed numerous inversions of the same synthetic Stokes profiles and compared their results with the original simulated atmosphere. This procedure has been carried out for two cases, one of them representing a quiescent atmosphere and the other during an umbral flash. In both cases, the inversions miss the actual vertical magnetic field but, interestingly, the error in the magnetic field is different. For the atmosphere at rest, the inferred field strengths are widespread and tend to underestimate the real value. During the umbral flash, the dispersion of the inversion solutions is lower, but most inversions overestimate the magnetic field strength. When observing a series of \CaII\ 8542 \AA\ Stokes profiles, this sequence of successive under- and over-estimations of the magnetic field strength can produce spurious field oscillations.  

Our results are restricted to inversions of the \CaII\ 8542 \AA\ line, but they point to the need to be cautious when interpreting magnetic field fluctuations inferred from the examination of the Stokes profiles from lines formed in NLTE. Recent works have shown that neglecting NLTE effects can lead to errors in the interpretation of the photospheric \FeI\ 6301.5 \AA\ and 6302.5 \AA\ lines \citep{Smitha+etal2020, Smitha+etal2021}. In the case of the chromospheric \CaII\ 8542 \AA\ line, we find that even employing NLTE inversions, the accuracy in the determination of the atmospheric quantities (especially the magnetic field) is limited. 

The analysis of observations with a fine spectral resolution, like those obtained from slit rather than imaging spectropolarimeters, provides a better estimation of the chromospheric magnetic field, in agreement with \citet{Felipe+EstebanPozuelo2019}. Imaging spectropolarimeters, like Fabry-P\'erot interferometers, can also introduce systematic errors in the estimation of the magnetic field due to the non-simultaneous acquisition of the different wavelengths of the spectral profile \citep{Settele+etal2002, Felipe+etal2018a}. However, even in the ideal case of high spectral resolution and instantaneous scan of the line, the error in the inference of the vertical magnetic field strength (as determined from the standard deviation of numerous good inversions) is up to 150 G. This result comes at no surprise since the \CaII\ 8542 \AA\ line has a relatively low sensitivity to the magnetic field. Thus, magnetic field variations must be critically evaluated and interpreted with care. Analyses of the solutions from independent inversion, as those illustrated in Figures \ref{fig:spread_rest}, \ref{fig:spread_UF}, \ref{fig:spread_rest_DKIST}, \ref{fig:spread_UF_DKIST}, are fundamental to ascertain the reliability of the results.

Studies can also benefit from independent inferences of the magnetic field. In the case of the \CaII\ 8542 \AA\ line such independent verification can be done with the weak field approximation. The applicability of the weak field approximation is restricted to those cases where the Doppler width of the line is larger than the Zeeman splitting produced by the magnetic field. The low magnetic sensitivity of the \CaII\ 8542 \AA\ line allows the use of the weak field approximation even in atmospheres permeated by a strong magnetic field, such as sunspot umbrae. Although the weak field approximation assumes the absence of vertical gradients in the magnetic field, it provides a simple and straightforward test for validation of inversion results. This approach was followed by \citet{delaCruz-Rodriguez+etal2013}, who found minor variations in the magnetic field strength between the quiescent and flash phases. 


Regarding the fluctuations in the chromospheric temperature and velocity, our study shows that they are appropriately recovered by the inversion of the \CaII\ 8542 line, and supports the use of this spectral line to investigate chromospheric oscillations in these parameters.  \citet{Keys+etal2021} has recently shown that inversions of the \FeI\ 6301 and 6302 \AA\ lines can recover the photospheric temperature, line-of-sight magnetic field, and line-of-sight velocity, providing a good characterization of the wave period. Their analysis was restricted to short-period waves, but the inversions managed to recover the small-scale fluctuations produced by the short wavelength of those oscillations. In upcoming work, we will examine the accuracy of the wave properties obtained from our inversions. This study will allow us to assess the potential of NLTE inversions to investigate wave propagation in the umbral chromosphere.

\acknowledgments

Financial support from the State Research Agency (AEI) of the Spanish Ministry of Science, Innovation and Universities (MCIU) and the European Regional Development Fund (FEDER) under grant with reference PGC2018-097611-A-I00 is gratefully acknowledged. We acknowledge the contribution of Teide High-Performance Computing facilities to the results of this research. TeideHPC facilities are provided by the Instituto Tecnol\'ogico y de Energ\'ias Renovables (ITER, SA). URL: \url{http://teidehpc.iter.es}.

\bibliography{biblio}{}
\bibliographystyle{aasjournal}



\end{document}